\documentclass[preprint]{aastex63}

\usepackage{amsmath}
\usepackage{natbib}
\usepackage{xcolor}
\usepackage{outlines}
\definecolor{darkgreen}{RGB}{0,127,0}

\newcommand{\grad}{\boldsymbol{\nabla}}
\newcommand{\cross}{\boldsymbol{\times}}

\newcommand{\del}[2]{\ensuremath{\frac{\partial #1}{\partial #2}}}

\providecommand{\e}[1]{\ensuremath{\times 10^{#1}}}

\received{\today}
\revised{\today}
\accepted{\today}
\submitjournal{ApJ}

\shorttitle{Reconnection Guide Field in Flares}
\shortauthors{Dahlin et al.}
\graphicspath{{./}{figures/}}

\begin{document}

\title{Variability of the Reconnection Guide Field in Solar Flares}
 
\correspondingauthor{Joel T.\ Dahlin}
\email{joeltdahlin@gmail.com, joel.t.dahlin@nasa.gov}

\author[0000-0002-9493-4730]{Joel T. Dahlin}

\author{Spiro K.\ Antiochos}
\affiliation{Heliophysics Science Division, NASA Goddard Space Flight Center,
Greenbelt, MD 20771}

\author{Jiong Qiu}
\affiliation{Department of Physics, Montana State University, Bozeman, MT, 59717}

\author{C.\ Richard DeVore}
\affiliation{Heliophysics Science Division, NASA Goddard Space Flight Center,
Greenbelt, MD 20771}

\begin{abstract}

Solar flares may be the best-known examples of the explosive conversion of magnetic energy into bulk motion, plasma heating, and particle acceleration via magnetic reconnection. The energy source for all flares is the highly sheared magnetic field of a filament channel above a polarity inversion line (PIL). During the flare, this shear field becomes the so-called reconnection guide field (i.e., the non-reconnecting component), which has been shown to play a major role in determining key properties of the reconnection including the efficiency of particle acceleration.  We present new high-resolution, three-dimensional, magnetohydrodynamics simulations that reveal the detailed evolution of the magnetic shear/guide field throughout an eruptive flare. The magnetic shear evolves in three distinct phases: shear first builds up in a narrow region about the PIL, then expands outward to form a thin vertical current sheet, and finally is transferred by flare reconnection into an arcade of sheared flare loops and an erupting flux rope. We demonstrate how the guide field may be inferred from observations of the sheared flare loops. Our results indicate that initially the guide field is larger by about a factor of 5 than the reconnecting component, but it weakens by more than an order of magnitude over the course of the flare. Instantaneously, the guide field also varies spatially over a similar range along the three-dimensional current sheet. {We discuss the implications of the remarkable variability of the guide field for the timing and localization of efficient particle acceleration in flares.}

\end{abstract}

\keywords{}

\section{Introduction} \label{sec:intro}


The solar atmosphere is replete with explosive activity, from ultraviolet (UV) bursts and Ellerman bombs at the smallest observed scales to spectacular X-class flares and associated coronal mass ejections that drive hazardous space weather. Many of these phenomena are thought to be due to the release of stored magnetic energy by the process of magnetic reconnection. In some remarkable events, much of the energy released is transferred to nonthermal particles, most notably in flares. The most extreme events apparently accelerate all of the coronal electrons \citep{krucker10a} and impart the majority of the energy to those energetic particles.
\par
Such highly efficient electron acceleration is not always observed in flares\citep{emslie12a,inglis14a,warmuth16a,aschwanden17a,warmuth20a}, and it is rare or difficult to observe in other instances of coronal reconnection. Two key questions naturally follow. Why is acceleration so efficient in flares? What factors control the considerable flare-to-flare variation of acceleration efficiency? We propose that the \textit{reconnection guide field} of the flare, in particular its evolution during the course of the flare, plays a critical role in the electron acceleration. In this article, we use a numerical model for an eruptive flare to investigate the origin of the guide field in filament channel shear and its subsequent evolution.
\par
It is well known that the highly sheared magnetic field of a filament channel is the energy source for coronal mass ejections and eruptive flares. The coronal magnetic field transitions to a lower-energy state by ejecting this shear in the form of a flux rope. Importantly, during the creation of the flux rope through flare reconnection, the shear participates in the flare energy release as the magnetic field's guide-field component.
The guide field is known to play a critical role in controlling the efficiency of plasmoid-associated particle acceleration \citep{dahlin16a,dahlin17a,li19b,arnold21a}. Recently, using a hybrid kinetic/magnetohydrodynamic (MHD) model, \cite{arnold21a} found that reconnection generates power-law nonthermal electron spectra that are highly sensitive to the ratio of the upstream guide field to the reconnecting field. Although the electron heating varied little with guide-field strength, the electron acceleration was suppressed in the regime where the guide field is much stronger than the reconnecting field. Therefore, the guide field appears to determine how efficiently flare-released energy is imparted to nonthermal electrons.

The guide magnetic field in flares is highly challenging to determine directly, due to the difficulty of measuring the coronal magnetic field. Nevertheless, several recent studies have indirectly estimated the guide field in flares.  \citet{chen20a} used Expanded Owens Valley Solar Array \citep[EOVSA;][]{gary13a} microwave observations to determine the magnetic-field profile along the current sheet during the 2017 September 10 X-class flare. They found best agreement with an ATHENA flare simulation assuming a guide-field ratio of 0.3. Using an indirect approach, \citet{qiu10a, qiu17a} estimated the guide field in a set of two-ribbon flares. They projected the apparent motion of newly brightened flare-ribbon fronts in the directions parallel and perpendicular to the polarity inversion line (PIL) and estimated the guide-field strength relative to the post-reconnection outflow field. The values ranged from 0.2 to 5.1 with a median value of about 1.

Although indirect measurements of the guide field are sparse, further information can be inferred from the evolution of the flare morphology and energy release. A frequent pattern observed in flares is the so-called ``strong-to-weak'' shear transition, wherein flare loops (or associated conjugate brightenings in UV or H$\alpha$) early in the flare tend to be highly sheared (nearly parallel) to the PIL, but the successively brightening loops are progressively less sheared as the flare evolves \citep{sakurai92a,aschwanden01a,asai03a,su06a,su07a,schmahl06a,aulanier12a}. This behavior also has been observed in conjugate hard X-ray footpoints \citep{masuda01a,bogachev05a,temmer07a,ji08a,liu09a_wei,yang09a}. Sheared structures imply the presence of a guide field at the reconnection site, and the strong-to-weak transition suggests a corresponding weakening of the guide field relative to the reconnecting component. Therefore, one of the objectives of this paper is to draw a quantitative link between the shear and the guide-field strength.
\par
In this article, we investigate the detailed guide-field evolution in new state-of-the-art, high-resolution simulations of an eruptive flare. Critically, we initiate these simulations from a potential-field configuration in order to model the fully self-consistent evolution of the flare, in particular the buildup of the shear that powers the eruption and becomes the guide magnetic field. In \S\ref{sec:theory}, we lay out the role of the shear/guide field in the standard model for eruptive flares. We present our numerical model in \S\ref{sec:model}. The evolution of the shear/guide field during the energy buildup and the eruption is presented in \S\ref{sec:flare}. We conclude by discussing the implications of our results for observations in \S\ref{sec:discussion}.

\section{Theoretical Background} \label{sec:theory}


The traditional CSHKP model for eruptive flares 
\citep{carmichael64a,sturrock66a,hirayama74a,kopp76a} interprets the primary morphological features, flare loops and ribbons, according to a reconnection paradigm. Flare loops are formed via reconnection in the corona, and the flare ribbons are illuminated by energy transport from the coronal reconnection sites to the lower levels of the solar atmosphere. This model has been highly successful in interpreting flare observations and has been corroborated by a number of 2.5D MHD simulations \citep[see][and references therein]{shibata11a}. The 2017 September 10 GOES X8.2 flare, in which the vertical current sheet was seen edge-on, strikingly resembled essential features of the CSHKP model including the erupting flux rope \citep{seaton18a, long18a, veronig18a, gopalswamy18a}, trailing current sheet \citep{seaton18a, warren18a, longcope18a}, and energy deposited by energetic electrons \citep{gary18a,chen20a}.

This standard model has been highly successful in interpreting flare observations, but it neglects several key elements present in actual events. First, the model assumes a simple, laminar current sheet, whereas both observations \citep{liu13a_w,kumar18a,cheng18a,kumar19a} and theory/modeling \citep{loureiro07a,bhattacharjee09a,karpen12a} have demonstrated that at large Lundquist numbers, reconnection generates numerous turbulent substructures known as plasmoids. Plasmoids are important for both particle acceleration \citep{drake06a,dahlin14a,li15a,guidoni16a} and fast energy release \citep{loureiro07a,bhattacharjee09a,karpen12a}. The role of plasmoids in flare dynamics and their impact on ribbon structure has recently been examined analytically by \cite{wyper21a}. We will discuss these issues further in a forthcoming paper.
\par
Another simplification is that the model essentially considers only the 2D reconnection of antiparallel fields at a magnetic null line (``X line''). In general, three-dimensional reconnection does not require a magnetic null, and there will be a magnetic-field component oriented along the X line and referred to as the guide field \citep{schindler88a,demoulin96a}. There is ample evidence for strong guide fields in the sheared filament channels and flare loops of solar eruptions. The guide field has recently been shown to control both the efficiency of particle acceleration by the Fermi mechanism in plasmoids \citep{dahlin16a,arnold21a} and the internal structure of the plasmoids themselves. For example, \citet{edmondson17a} found that the correlation length of magnetic fluctuations along the guide-field direction increases with the guide-field ratio.

There have been a number of efforts to extend the CSHKP model to 3D configurations, in particular to establish interpretations for features not captured in the 2D picture, including sigmoids, sheared flare loops, and ribbon morphologies and locations. Using a 3-D numerical model of an eruptive flare triggered by torus instability, \citet{aulanier12a} found successively reconnected loops diminished in shear, consistent with the observational picture. This was attributed primarily to the spatial distribution of shear in the pre-eruptive field, which was strongest near the PIL, but also to  
lengthening of the CME flux-rope field lines prior to flare-reconnection onset that thereby diminished the shear strength.

The evolution of the shear/guide field in an eruptive flare occurs in three phases. It is well-known that CMEs/eruptive flares originate from highly sheared filament channels \citep{gaizauskas98a,gaizauskas01a,martin98a}, where the magnetic field is oriented nearly parallel to the polarity inversion line (PIL). The degree to which these filament channels contain magnetic twist (i.e., a flux rope) has seen significant debate \citep{forbes06a,patsourakos20a}. Nevertheless, the shear-field component is ultimately the energy source for the eruptive flare, and much of the energy release is due to the ejection of the shear field into the heliosphere. The shear contributes excess magnetic pressure that expands the field outward, stressing the system. Regardless of the onset mechanism, in order for significant energy release to occur via the flare reconnection, this energy must be transferred from the shear component to the radial component that is processed by the flare. This occurs primarily by the shear field stretching out the radial field to form a narrow current sheet.  When this current sheet reconnects, magnetic flux carrying shear is processed by the flare: the flux ejected upward becomes the toroidal component of the erupting flux rope, and that carried downward becomes the shear in the flare loops. The shear flux, however, is localized to the filament channel, and it is eventually exhausted as the flare reconnection progresses to process unsheared flux outside of the filament channel \citep[e.g.][]{su06a}. Therefore, the reconnection guide field must weaken substantially over the course of the flare.

We conclude from these arguments that the reconnection guide field is simply the sheared field of the filament channel. An accurate flare model must include the self-consistent response of the coronal magnetic field to the shear buildup and the subsequent creation of the flare current sheet. The details of the magnetic configuration upstream of the current sheet determine where and when the flare begins. In order to investigate the origin and evolution of the guide field, therefore, it is critical to model self-consistently the gradual energy buildup prior to the explosive flare onset.

\section{Numerical Model} \label{sec:model}

\begin{figure}[ht!]
\plotone{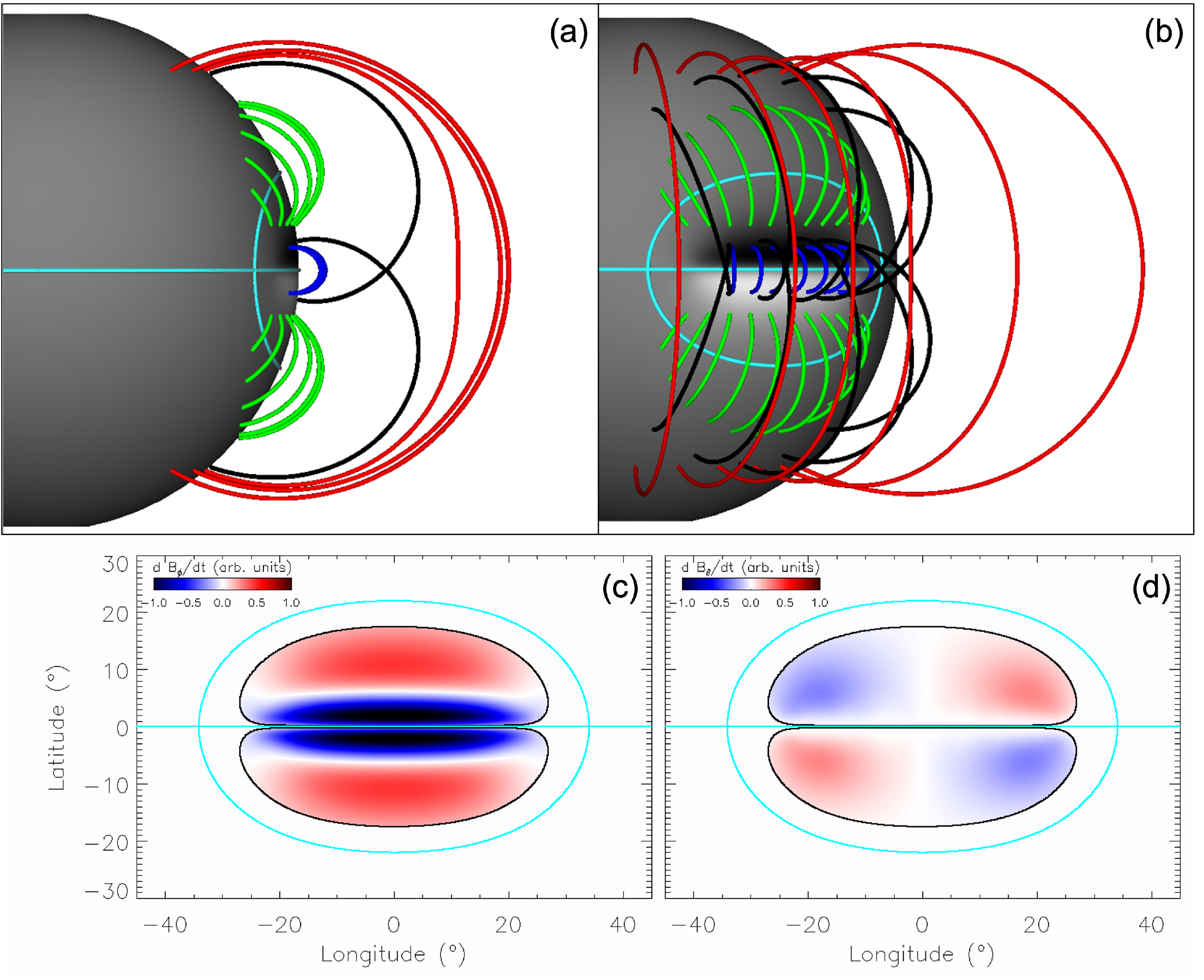}
\caption{(a,b) Initial magnetic configuration for the eruptive flare simulation. White (black) shading at the inner boundary corresponds to positive (negative) polarity. Polarity inversion lines (PILs) are indicated by cyan curves. Four distinct flux regions are indicated in blue, green, and red (northern and southern sets of green field lines represent distinct flux regions). (c,d) Profile of tangential flux injected by the STITCH method (see text for details) at the inner radial boundary, normalized to 0.58 G s$^{-1}$. The $B_\phi$ component is shown in (c) and the $B_\theta$ component in (d).
\label{fig:config}}
\end{figure}

Our eruptive flare simulation was performed with the
Adaptively Refined Magnetohydrodynamics Solver
\citep[ARMS;][]{devore08a}, which has been used to model both
CMEs/eruptive flares \citep[see
also][]{lynch08a,lynch09a,karpen12a,masson13a,lynch16a,masson19a} and the formation
of filament channels
\citep{zhao15a,knizhnik15a,knizhnik17a,knizhnik17b,knizhnik18a,dahlin19a}.
The magnetic-field configuration (illustrated in Fig.~\ref{fig:config}a,b) is an idealized active region consisting of two sets of dipoles located just beneath the solar surface at the equator, superposed on a background solar dipole (see Appendix). This combination generates four distinct flux systems in a multipolar ``breakout'' topology \citep{antiochos98a,antiochos99a}. The antisymmetric magnetic-field profile across the equator generates a null line above the active region that intersects with the solar surface where the PILs cross east and west of the active region. This symmetry is quickly broken by the energy injection as described below. The configuration was chosen so that the flare current sheet would align nearly perfectly with the equator, to simplify the analysis of the shear/guide-field evolution.

The initial atmosphere was a spherically symmetric
hydrostatic equilibrium with an inverse-$r$ temperature profile
at base temperature $T_s = 2 \times 10^6$ K and pressure
$P_s = 4 \times 10^{-1}$ dyn cm$^{-2}$ \citep[as in][]{dahlin19a}. 
We solved the ideal MHD equations with an adiabatic
temperature equation \citep[as in][]{devore08a,karpen12a,dahlin19a}. The domain
extents were $r \in [1R_s, 30R_s]$, $\theta \in [\pi/16, 15\pi/16]$,
and $\phi \in [-\pi,+\pi]$, where $R_s$ is the solar radius. Grid-scale numerical dissipation breaks the frozen-in flux constraint enabling reconnection to occur.
\par

We used the adaptive-mesh capability of ARMS to resolve selectively important fine-scale structure. The refinement criteria are based on the ratios
of the local spatial scale of the electric-current
density to the local grid spacing, as in previous high-resolution calculations \citep{karpen12a,masson13a,masson19a}. The base (level 1) block structure is $7 \times 7 \times 16$ in $r$, $\theta$, and $\phi$, respectively, and each block contains $8 \times 8 \times 8$ grid cells. For our highest-resolution calculations, a minimum refinement level of 6 is imposed at the innermost boundary over the entire active region ($\theta \in [3\pi/8, 5\pi/8]$, $\phi \in [-\pi/4, \pi/4]$) and a level of 4 over the same angular region for $r \in [1 R_s, 2 R_s]$.
A refinement level of $n$ corresponds to $2^{n-1}$ times the resolution of the base block structure at level 1. To reduce the computational expense, the overall maximum refinement was held at level 6 during the early energy injection/filament channel stage of the evolution, $t \in [$0~s,11500~s$]$. Subsequently, the adaptive mesh was allowed to refine up to a maximum level of 8. 
Adaptive refinement/derefinement is restricted to the region $B>$~2~G, so that the enhanced resolution is not applied to the breakout current sheet. 
The smallest grid cell size occurs at the inner boundary, $\Delta r = 0.33$~Mm and 
$R_s \Delta \theta = R_s \Delta \phi = 0.27$~Mm.


We inject shear flux to form the filament channel using the method of statistical injection of condensed helicity \citep[STITCH;][]{mackay14a,mackay18a,lynch21a,dahlin21a}. The injection of shear flux at the inner boundary is given by
\begin{equation}
\label{eqn:stitch_expression}
\del{\mathbf{B_\perp}}{t} = h^{-1} \grad_\perp \cross \left(\zeta_s B_r \hat{\mathbf{r}}\right),
\end{equation}
where $\mathbf{B_\perp}$ is the tangential magnetic field, $B_r$ the radial component, $\grad_\perp$ the tangential derivative, and $h$ the height of the flux-injection region (one grid cell). The parameter $\zeta_s$ incorporates the local rate of rotation (i.e., vorticity) of the helicity-injecting surface flows \citep{antiochos13a,mackay14a} and reconnection of the induced twist flux at current sheets, all of which occur at sub-grid scales in this simulation. The spatial and temporal profiles of these processes are specified by 
\begin{equation}
\zeta_s(\xi,\phi,t) = \zeta_0 f(t) \sin \left[\frac{\pi\xi}{\xi_0}\right] \sin \left[\frac{\pi\phi}{\phi_0} \right],
\end{equation}
where $\zeta_0 = 4.0\times 10^{15}$~cm$^2$~s$^{-1}$,\, $f(t)$ is the temporal profile, $\xi = \pi/2-\theta $ is the latitude coordinate, 
$\xi_0 = 0.125\pi$ $(22.5^\circ)$, and $\phi_0 = 0.195\pi$ $(35.1^\circ)$. In addition, we set $\zeta_s = 0$ where $|B| < 3.5$~G within each semi-elliptical flux region, as well as outside of the central pair of polarities that comprise the idealized active region. The instantaneous profiles for the injection of tangential flux are shown in Figure \ref{fig:config}c,d. 
The flux injection is proportional to $f(t)$, which is ramped up from zero initially, held fixed at unity for an extended time, then is ramped back down to zero prior to the eruptive flare. This temporal profile is shown as the magenta curve in Figure~\ref{fig:energy} below.
The magnetic field is line-tied at rest, with all velocity components fixed at zero, across the entire inner radial boundary. Otherwise, we apply the same boundary conditions used in our previous 3D ARMS eruptive-flare calculation \citep[][]{dahlin19a}.
\par



\section{Results} \label{sec:flare}

\begin{figure}[ht!]
\plotone{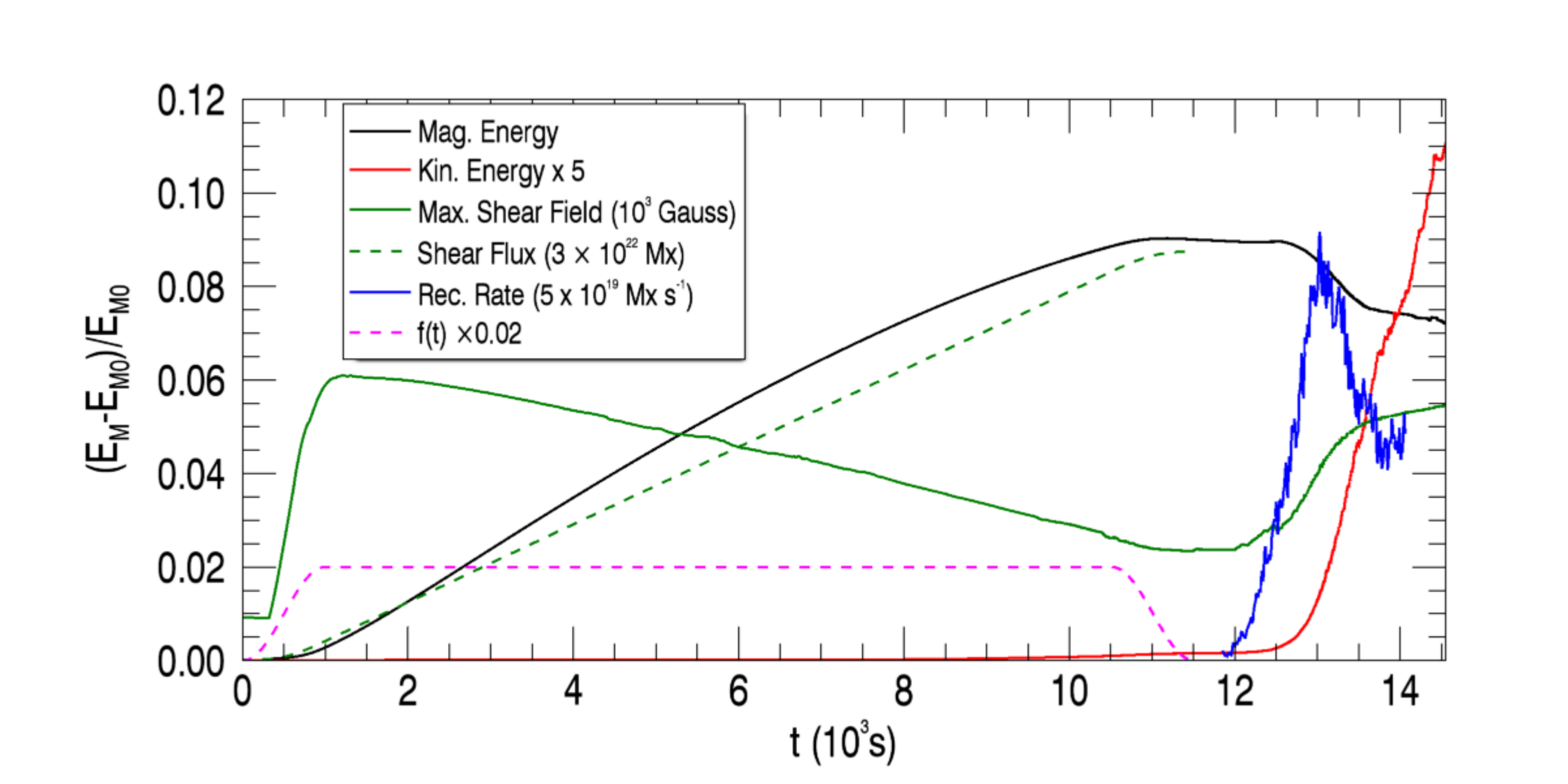}
\caption{Global energetics during the energy buildup and the eruptive flare. The magenta curve indicates the temporal profile of shear injection, solid (dashed) green the maximum strength (total flux) of the shear field, black the magnetic free energy, red the kinetic energy, and blue the rate of reconnected flux. $E_M$ denotes the total magnetic energy and $E_{M0}$ its initial value.
\label{fig:energy}}
\end{figure}

 Shear flux was injected gradually at the inner boundary using the STITCH method until the system erupted explosively. Through an iterative process, we identified how much injected flux was sufficient to drive the system to eruption and halted the STITCH injection just at this point. The system thereafter evolved quasi-statically, primarily due to breakout reconnection, prior to the onset of the fast dynamics. The Alfv\'en speed is $c_A \approx 2000$~km~s$^{-1}$ and the length scale of the active region is $L = 400$~Mm, so the characteristic Alfv\'en propagation time $t_A \approx 200$~s. This is significantly smaller than the time between the halting of the driving and the onset of eruption, as shown in Figure \ref{fig:energy}.
\par

The temporal evolution of several global quantities characterizing the eruptive flare is shown in Figure~\ref{fig:energy}. The energy-buildup phase extends over 0~s $< t <$ 11,500~s, whereas the eruptive-flare onset occurs at $t \approx 12,000$~s. The globally integrated magnetic free energy (solid black curve) closely follows the total shear flux in the filament channel (dashed green curve). Similar to our findings in a previous eruptive-flare study \citep{dahlin19a}, the maximum shear field (solid green curve) rises rapidly to a maximum during the energization phase, then decreases gradually thereafter until flare onset. This behavior is due to the increase in the filament channel's cross-sectional area as the flux expands upward and outward in response to the buildup of shear magnetic pressure. Such evolution is predicted by the Aly-Sturrock limit \citep{aly84a,sturrock91a}, which argues that in both the minimum- (closed potential) and maximum- (fully open) energy states, the shear field goes to zero. Hence, the shear-field strength must peak intermediate to those energy extrema.
\par

The reconnection rate ($d\Phi_{rec}/dt$, where $\Phi_{rec}$ is the reconnected flux) reveals the rapid onset and fast dynamics of the eruptive flare. The rate is determined by high-cadence tracking of changes in the magnetic-field connectivity within the erupting active region \citep[as in][]{dahlin19a}. This process will be discussed in depth in a forthcoming study that focuses on flare-ribbon fine structure. The onset of fast reconnection precedes the rapid conversion of magnetic energy into kinetic energy, as has been noted in previous studies \citep{karpen12a,dahlin19a}. The maximum shear-field strength increases again at flare onset, due to the retraction and compression of flare loops that contain some shear flux. However, the majority (approximately $80\%$) of the filament-channel shear is ejected by the eruption. 
The magnetic free-energy release is much smaller because the erupting flux rope containing most of the shear remains within the simulation domain.
\par
    
\begin{figure}[ht!]
\plotone{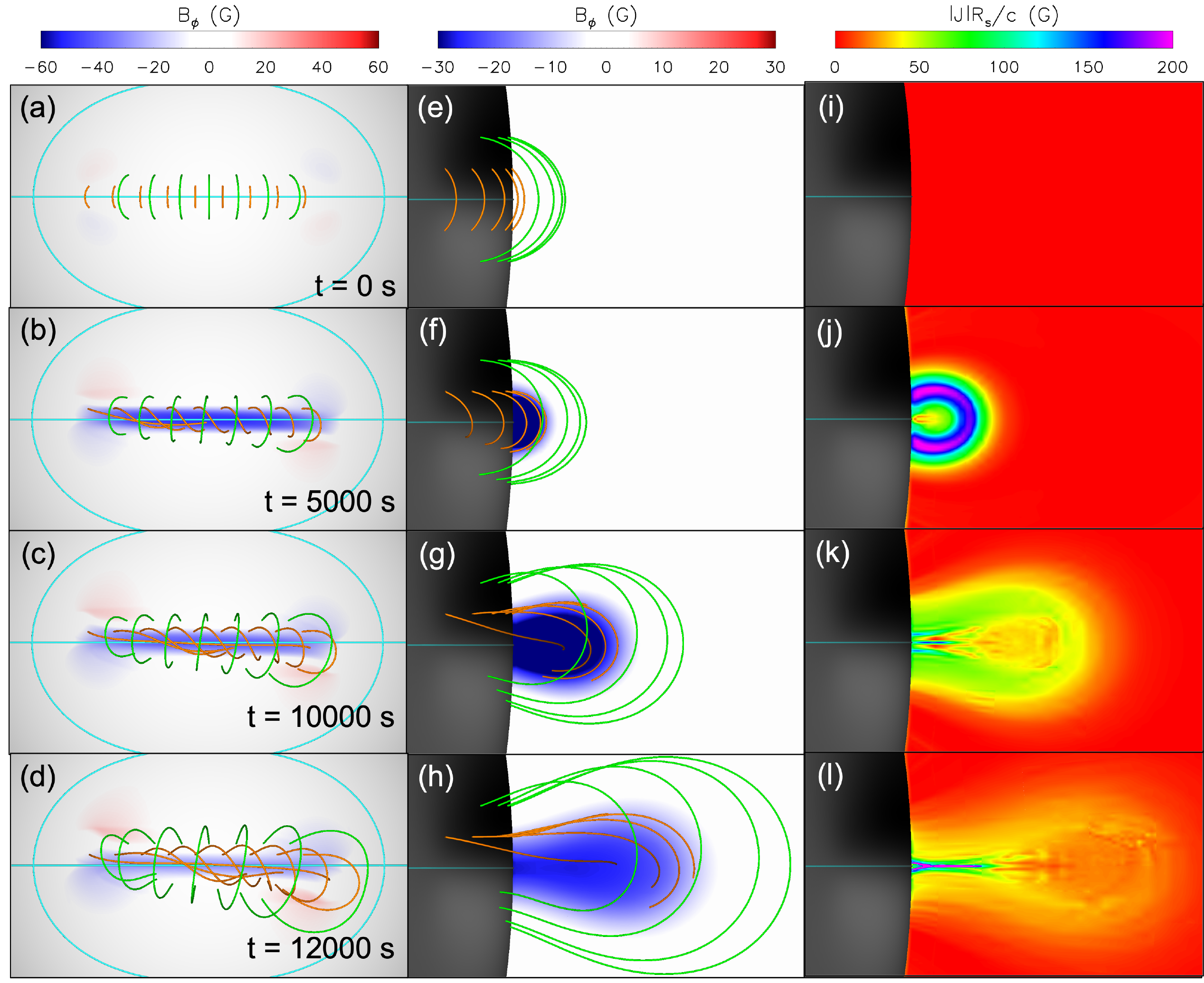}
\caption{Energy buildup phase of the eruptive flare. (a-d) Overhead view of the filament channel, with two sets of magnetic field lines corresponding to the low-lying (orange) filament channel and overlying (green) arcade loops. Blue-to-red shading indicates $B_\phi$ at the inner boundary. (e-h) $\phi = 0$ cut of $B_\phi$ (shear/guide field), with the same sets of field lines shown in (a-d). (i-l) $\phi= 0$ cut of $\left\vert \mathbf{J} \right\vert R_s/c$ (normalized current-density magnitude,  where $R_s$ is the solar radius).
\label{fig:shear}}
\end{figure}

The formation and evolution of the filament channel are illustrated in Figure \ref{fig:shear}, which shows representative filament-channel field lines along with $\phi = 0$ cuts of $B_\phi$ and normalized current-density magnitude $\left\vert \mathbf{J} \right\vert R_{s}/c$. The magnetic field is initially potential (current-free), so the field lines run perpendicular to the PIL forming an unsheared arcade. As shear flux is injected into the system, the low-lying orange field lines become stretched along the $\phi$ direction. The red field lines are more weakly sheared, and they expand upward and sideways. Note that for weak shear the overlying flux expands negligibly so the distribution of the shear flux (Fig.~\ref{fig:shear}f) reflects directly the gradients in the shear injection by STITCH. This gives rise to the strong current shell observable in Figure ~\ref{fig:shear}j, coincident with the boundary of the sheared region. As the driving continues, the shear flux expands outward, eventually assuming an inverted teardrop shape (Fig.~\ref{fig:shear}g).The expansion causes a decrease in the magnitude of the shear field, Figure~\ref{fig:energy}, so the current shell disappears, but a new vertical current sheet forms deep within the shear region (Fig.~\ref{fig:shear}k). This current sheet forms only quite late in the evolution, after $t = 10,000$~s and shortly before the driving is halted. Figure~\ref{fig:shear} also shows that the region of strong shear field near the PIL is \underline{not} identical to the region of strong shear injection; compare panels (f) and (h). The sheared field lines farther from the PIL expand more freely upward and outward, distributing the shear flux over a far greater volume than would be predicted from injection into the initial potential field. The region of strong shear that is relevant for the flare-reconnection guide field is, therefore, much narrower than the full region over which the shear is injected.
\par
     
\begin{figure}[ht!]
\plotone{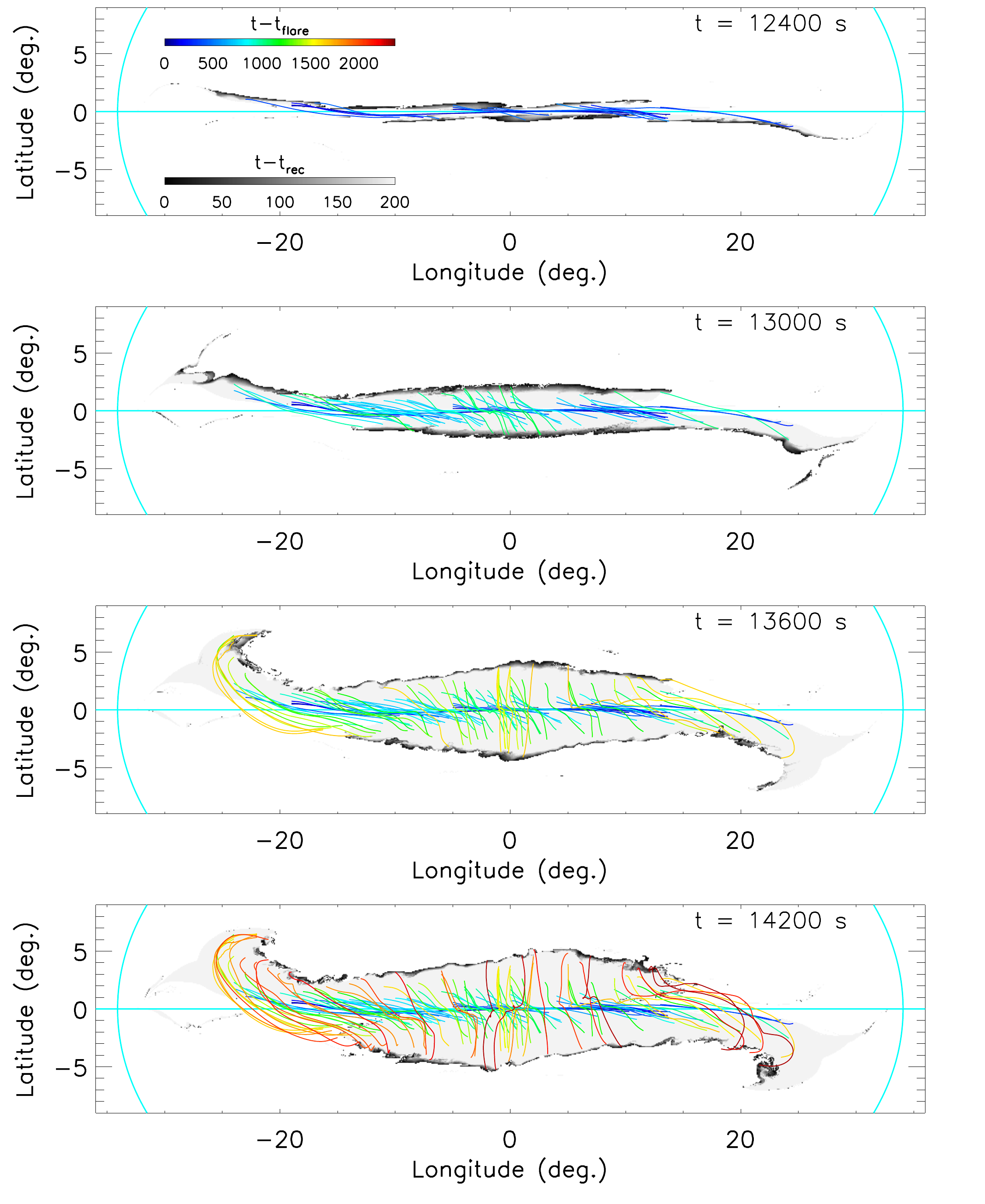}
\caption{Flare evolution illustrated by flare ribbons and loops. Ribbons are represented by the footpoints of reconnected field lines, colored according to the elapsed time ($t-t_{\rm{rec}}$; gray scale) since the reconnection occurred on that field line ($t_{rec}$). Flare loops are represented by reconnected field lines, colored according to the time ($t-t_{\rm{flare}}$; color scale) since flare onset $t_{flare} = 11850$ s. 
An animation of this figure is {available, showing the evolution of the flare ribbons and loops at $10$~s cadence for $11860$~s $\leq t \leq 14220$~s (the animation duration is 16 seconds). Note the grayscale color map for the flare ribbons is reversed in the animation.}
\label{fig:ribbons_loops}}
\end{figure}

 The flare current sheet (Fig.~\ref{fig:shear}l) lengthens and its aspect ratio increases until it eventually 
 transitions to fast reconnection. The ensuing flare reconnection is illustrated in Figure~\ref{fig:ribbons_loops} by tracing field lines from a grid of $901\times226$ footpoints at the inner boundary over the interval 11,850~s $< t <$ 14,520~s. We identify a reconnection event as occurring at the time when the field line traced from a given footpoint shortens permanently by at least $40\%$ relative to its maximum value. Figure \ref{fig:ribbons_loops}a shows that initially the ribbons of opposite magnetic polarity are displaced in longitude and are joined by loops that are highly sheared and run nearly parallel to the PIL. As the reconnection proceeds, successively formed loops are progressively less sheared, until eventually the reconnected loops in the central region run nearly perpendicular to the PIL. This evolution is consistent with the strong-to-weak shear transition long observed in flares. 
 We point out that the ``ribbons'' and ``loops'' discussed here are derived solely from the magnetic-field evolution and represent diagnostics of the reconnection, rather than forward modeling of the observable analogues. The latter would require a much more detailed treatment of the plasma thermodynamics and radiation. For example, the flare loops would typically be observed in EUV after a substantial time delay with respect to the corresponding H$\alpha$ ribbons.
\par

 \begin{figure}[ht!] \plotone{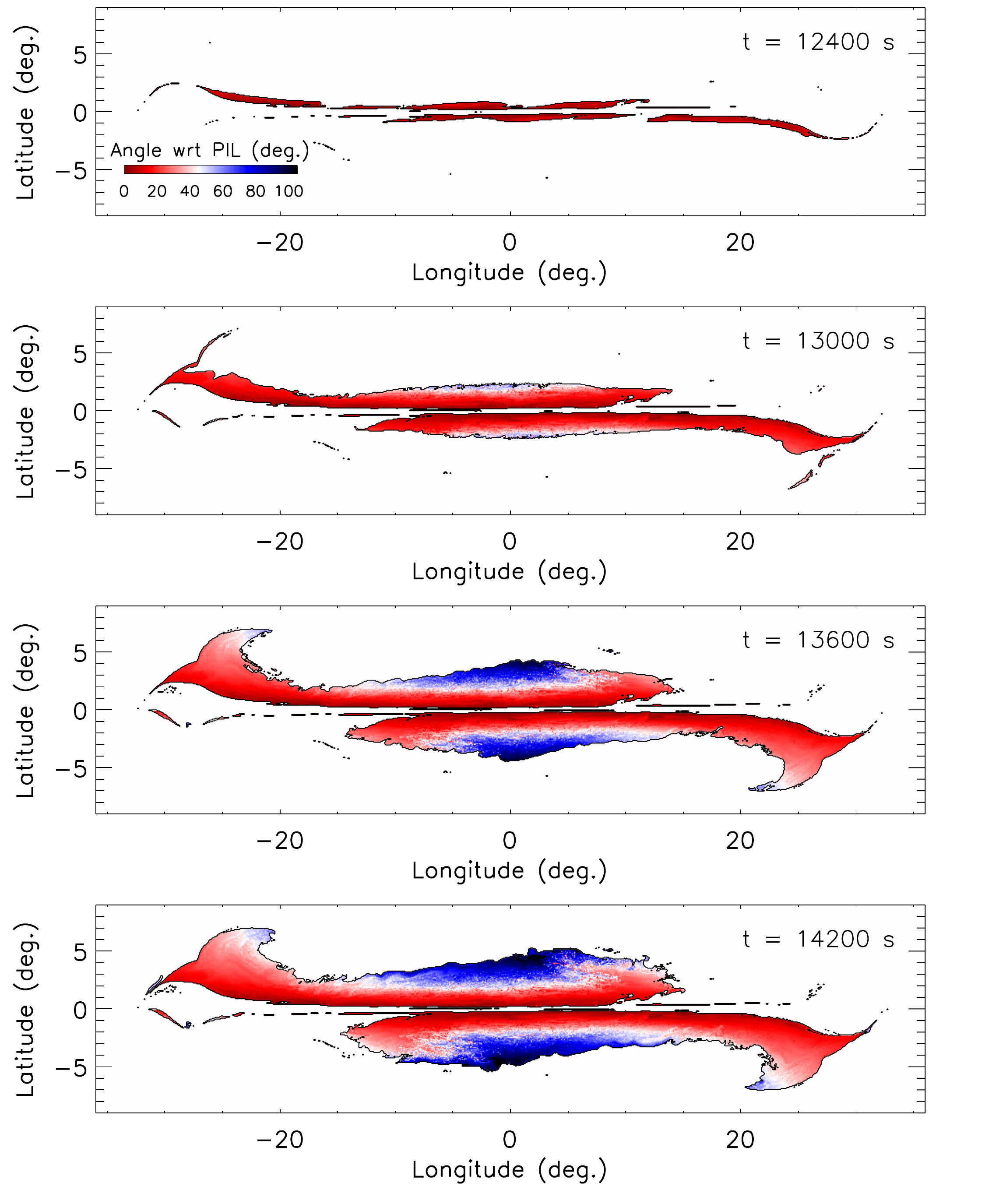} \caption{Cumulative flare ribbon kernel at several time steps during the evolution of the eruptive flare. Each footpoint is colorized according to the angle $\alpha$ that the corresponding flare loop makes with respect to the PIL, calculated according to its conjugate footpoints. 
 An animation of this figure is {available, showing the evolution of the cumulative flare ribbon kernel at $10$~s cadence for $11860$~s $\leq t \leq 14550$~s (the animation duration is 18 seconds).}
 \label{fig:ang_ribbons}} \end{figure}

Figure~\ref{fig:ribbons_loops} also shows the characteristic ``hook'' or ``J-shape'' at the ends of the ribbons. These are well-known and have been studied extensively in both observations \citep[e.g.,][]{Yuwei20} and simulations \citep[e.g.,][]{Janvier14}. They are due to presence of the footpoints of the escaping CME flux rope at the centers of the hooks. In 3D the ribbons must end at the CME footpoints, as illustrated very clearly in Figure 7 in \citet{Janvier14}. Our results agree very well with this large body of work on ribbon shape.

In order to examine the shear evolution quantitatively, we calculate the shear angle $\alpha$ for all of the flare loops identified via field-line tracing (Fig.~\ref{fig:ang_ribbons}). The angle between the flare loop and the PIL is defined by $\alpha = \arctan (\Delta \theta/\Delta \phi)$, where the displacements of the conjugate footpoint positions are $\Delta \theta$ and $\Delta \phi$. During the initial stage of reconnection (Fig.~\ref{fig:ang_ribbons}a,b), all flare loops are highly sheared with respect to the PIL ($\alpha < 45^\circ$). As reconnection proceeds (Fig.~\ref{fig:ang_ribbons}c,d), the decreasing shear of the central flare loops is readily apparent. However, we find a substantial variation in shear along the ribbons.
Although the flare loops near the central region of the instantaneous ribbon are largely unsheared at late times (Fig.~\ref{fig:ang_ribbons}c,d), the loops near the ends of the ribbons still have substantial shear.  Note also that the width of the strong-shear region ($\alpha < 45^\circ$) is much narrower than might be expected from the STITCH profile (Fig.~\ref{fig:config}c), which injects shear out to $\pm 5^\circ$ in latitude from the PIL.  We clarify the physical origins of these results in the Discussion below.

Thus far, we have focused on the evolution of the magnetic shear, which has two important characteristics: (a) an overall evolution from strong-to-weak shear, consistent with observations; and (b) substantial instantaneous variations along the flare ribbons. As previously noted, the shear in the flare loops is a clear indicator of the presence of a guide-field component at the originating reconnection site. However, 
quantitatively relating the measured loop shear 
to the upstream guide-field ratio at the flare current sheet is not simple. The guide-field parameter described in particle-acceleration simulations \citep[e.g.][] {dahlin16a,li17a,arnold21a} refers to the upstream (relative to the reconnection site) ratio of axial/out-of-plane component to the in-plane/reconnecting component of magnetic field. In an eruptive flare model, this is challenging to calculate as the location of the current sheet evolves over time. Furthermore, in contrast to the kinetic simulations where the upstream magnetic field approaches a uniform value, there is large-scale variation in the upstream magnetic field in the eruptive-flare model. In addition, in our simulation there is the added complexity of fully 3D structure and dynamics.
\par

\begin{figure}[ht!]
\plotone{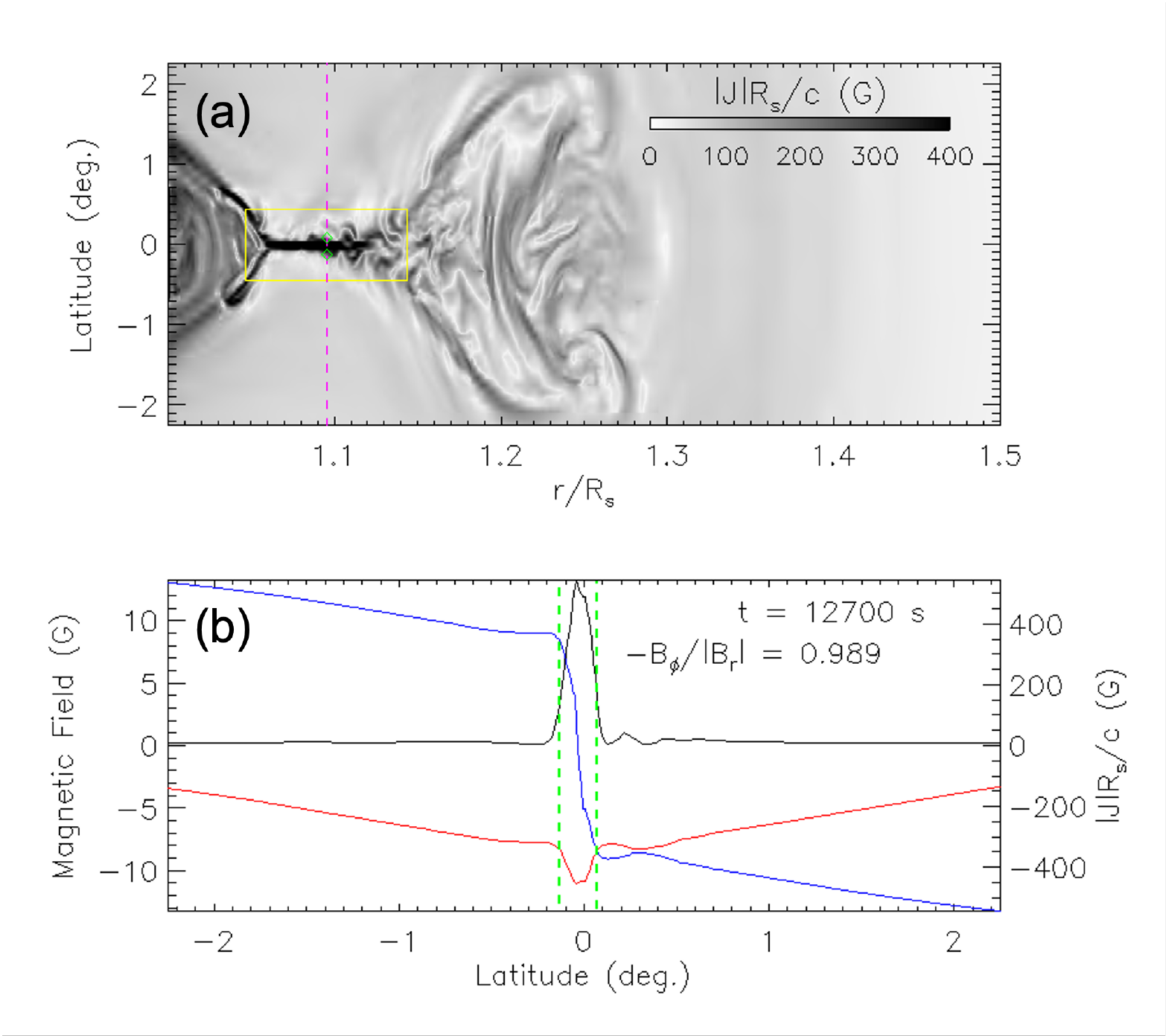}
\caption{Illustration of the method for identifying the reconnection guide field. (a) Normalized current-density magnitude. The yellow box indicates the automatically identified flare current-sheet region. The dashed magenta line is a cut through the middle of the sheet. (b) Profiles of the reconnecting magnetic field component ($B_r$, blue), guide-field component ($B_\phi$, red), and normalized current-density magnitude (black) along the cut shown in (a). Green dashed lines show where the upstream normalized guide field $b = -B_\phi/|B_r|$ is calculated, marked by green diamonds in (a). {An animation of this figure is available, showing the evolution of both the current sheet density (a) and the magnetic field profile (b) at $10$~s cadence for $11860$~s $\leq t \leq 14550$~s (the animation duration is 18 seconds).}
\label{fig:bgfind_alt}}
\end{figure}

We devised an algorithm to estimate the guide-field ratio, illustrated for one time step in Figure~\ref{fig:bgfind_alt}. First, it is necessary to identify the location of the current sheet; we examine the guide field in the center of the configuration at $\phi = 0$. At the ends of the flare current sheet (Fig.~\ref{fig:bgfind_alt}a), the current density bifurcates into the flare loops below and into the outgoing flux rope above. To first order, the flare current sheet can be identified as the region where the current density is single-peaked. However, the current sheet also can bifurcate internally due to transient plasmoids. Hence, we require that the current density reach its peak within a prescribed window near the equator, and we allow this window to broaden linearly in time at an empirically chosen expansion rate. The flare current sheet is taken to be the longest contiguous region that contains all of the current-density peaks within the specified window. The current density and the yellow box indicate the identified current-sheet region at $t = 12,700$~s
(Fig.~\ref{fig:bgfind_alt}a).
\par

To calculate the guide-field ratio, we take a constant-radius cut (dashed magenta line) through the center of the identified region. We fix the ``upstream'' locations as the points, approaching the current sheet from either direction, where the current density first reaches $25\%$ of its peak value (green diamonds in Fig.~\ref{fig:bgfind_alt}a, dashed lines in Fig.~\ref{fig:bgfind_alt}b). The $25\%$ value was chosen empirically: we found that this choice reasonably identified upstream regions throughout the entire evolution. As Figure \ref{fig:bgfind_alt}b shows, the green dashed lines match well with the inflection points of both the guide-field (red) and the reconnecting-field (blue) components. At $t = 12,700$~s, the time shown in Figure \ref{fig:bgfind_alt}, the guide-field ratio averaged from both sides of the current sheet is approximately unity.
\par

\begin{figure}[ht!]
\plotone{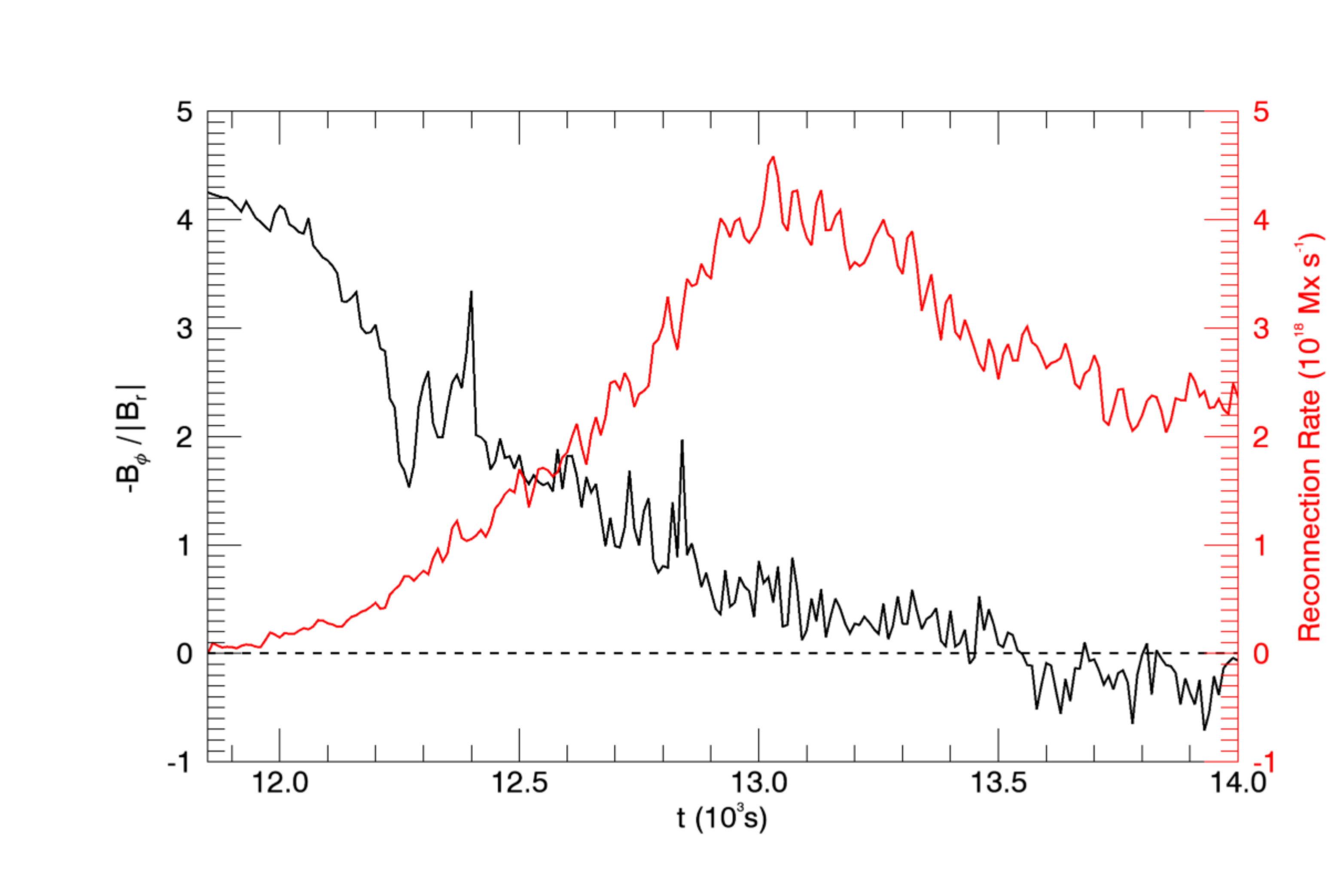}
\caption{
Guide-field ratio (black) and rate of total reconnected flux (red). During the phase of fast reconnection, the guide-field ratio $b$ weakens from $b \approx 4$ to $b \approx 0$, with a value of $b \approx 0.5$ at the time of peak reconnection rate.
\label{fig:bgplot}}
\end{figure}

We calculate the guide-field ratio in this manner over the entire interval 11,850~s $< t <$ 14,000~s. The resulting guide-field ratio is plotted in Figure~\ref{fig:bgplot} against the reconnection rate, $d\phi_\text{rec}/dt$. 
In the convention adopted here, the ratio can be negative; we include the sign of $B_\phi$ in order to clarify when the guide field passes through zero. Early in the flare, the guide field is strong with magnitude exceeding unity. In this regime, the particle acceleration would be suppressed, according to recent kinetic results \citep{dahlin16a,arnold21a}. 
Thereafter, the guide field weakens to zero, even reversing its direction by the time the flare reconnection rate peaks. (This reversal occurs due to a zone of weak, reversed shear away from the PIL, evidenced by the slight opposed tilt of overlying arcade loops (red) in Fig.~\ref{fig:shear}b-d.) Therefore, we conclude that the guide field (a) is large enough to suppress particle acceleration in the early phase of fast reconnection but (b) weakens rapidly to enable particle acceleration in the late phase. These two distinct phases are likely to characterize all cases of fast flare reconnection: an early, strong guide-field phase, which primarily drives heating; and a late, weak guide-field phase, which results in efficient particle acceleration.
\par

\section{Discussion}\label{sec:discussion}



The results presented in this paper demonstrate that a substantial and significantly variable guide field is an inevitable feature of flare reconnection. 
The guide field is represented by the shear component of the magnetic field within the flaring filament channel, i.e., the component aligned with the polarity inversion line. 
Our high-resolution simulation quantifies the detailed spatial and temporal variability of the guide field throughout the flare evolution. Although the present simulation used the ``breakout'' mechanism \citep{antiochos98a,antiochos99a}
for eruption onset, the essential features of the flare reconnection should hold for all onset mechanisms including, e.g., the torus instability \citep[cf.][]{aulanier12a}. This broad conclusion follows from a straightforward consideration of the generic eruptive-flare process.
\par

The source of free energy for all CMEs/eruptive flares is the strong shear in the filament-channel magnetic field; the twist component, if any, is small by comparison prior to eruption. 
The eruption ejects the shear by the following process.  First, a sheared filament-channel field line greatly increases its length by expanding upward and, eventually, becoming part of 
the CME flux rope (Fig.~\ref{fig:shear}h). 
In a force-free field, the shear component tends to distribute itself uniformly along the field-line length \citep{parker79a}. Consequently, the amount of shear that remains in the flare loops, versus the amount that is ejected in the CME flux rope, is determined simply by the location of the reconnection along the loop. Early on, when the CME flux rope is first forming and the reconnection occurs near the top of the innermost field lines, most of the shear remains behind in the initially formed flare loops. As the reconnection progresses, on the other hand, the CME flux rope expands and rises rapidly, so that an increasing share of the shear is above the reconnection point and is ejected rather than being left at the Sun. The expansion of the CME flux rope is highly nonuniform in 3D, however. It is fastest and largest in the central region of the ejecta, and significantly less so in the end regions. 
Consequently, the ``outer'' flare loops that 
close over the end regions retain more shear than the ``inner'' flare loops that 
close over the central region. 
The net result is that there is: an overall strong-to-weak shear transition in time; 
a highly variable shear along the instantaneous flare ribbons; and 
a post-eruption residual shear region that is much narrower than the initial filament channel.  \par

An important question raised by the work above is whether the location within the current sheet where the reconnection occurs can be determined from observations of chromospheric ribbons and the footpoint shear.  Unfortunately, this does not appear to be straightforward. If the field remains roughly force-free throughout its evolution, then the shear will be uniformly distributed along a field line, so the ratio of the shear at a footpoint before and after flare reconnection is equal to the ratio of the location of reconnection along that line to its total length. Therefore, if the pre- and post-flare footpoint shear could be measured accurately, this would constrain the location of reconnection. To determine exactly where the reconnection occurred, however, would require knowing the total length of the field line at the instant of reconnection. To our knowledge, this information is not readily available from present observations. \par 

Our results have two important implications for understanding flare energy release. First, recent studies demonstrate that plasmoid-driven particle acceleration is suppressed by a strong guide field \citep{dahlin16a,arnold21a}. In contrast, bulk plasma heating is largely insensitive to the guide field. We expect, therefore, that plasma heating should dominate over particle acceleration in the early phase of flares when the guide field is strong. This may explain so-called ``hot onset'' events in which strong heating is observed to precede significant particle acceleration \citep{hudson21a}. 
Furthermore, strong guide fields enhance the escape of electrons from plasmoids  \citep{dahlin17a,dahlin20a}. This allows the particles to be accelerated repeatedly by newly forming plasmoids in adjacent regions of the flare current sheet \citep{dahlin15a,dahlin17a,li19b,zhang21a}. These processes are expected to delay further the appearance of signatures of non-thermal electrons accelerated by the flare.
\par

Second, substantial spatial variation of the magnetic shear, in particular the persistence of strong shear near the flux-rope footpoints, implies similarly substantial variability in the guide magnetic field. 
The ensuing strong variations in the acceleration of energetic flare particles should manifest in the morphologies of hard X-ray footpoint and loop-top sources at the Sun, as well as the morphologies of prompt solar energetic particle events that arise from rapid escape of flare-accelerated particles into the heliosphere. 
In this paper, we simulated a highly idealized system with smooth large-scale structure, rather than an observed configuration with complex active-region structure at small scales. A more realistic setup could lead to even greater localization of both strong and weak guide-field regions, which could impart similar structure to the flare HXR sources. Despite these limitations, the work presented here provides a great deal of insight into understanding solar observations, through our detailed analysis of the variability of the guide magnetic field during flare reconnection.
\par

\acknowledgments

This work was supported by NASA via the SOLFER DRIVE Center at the University of Maryland, College Park, the H-LWS and H-ISFM programs at NASA Goddard Space Flight Center, and the H-GI program at Montana State University. J.T.D.\ 
was supported by an appointment to the NASA Postdoctoral Program at NASA GSFC, administered by Universities Space Research Association under contract with NASA.

\appendix

The initial magnetic field configuration is given by the superposition of thirty-one magnetic dipoles placed beneath the inner boundary. The total magnetic field is given by
$\mathbf{B}(\mathbf{r}) = \sum \mathbf{B}_{i}^{\text{dip}}(\mathbf{r})$, where 

\begin{equation*}
\mathbf{B}_i^{\text{dip}} = M_i\left(\frac{R_{i}^{\text{dip}}}{|\mathbf{r}-\mathbf{r}_{0,i}}\right)^3 [3\mathbf{n_i}(\mathbf{n_i} \boldsymbol{\cdot} \mathbf{m_i}) - \mathbf{m_i}],
\end{equation*}
$\mathbf{n}_i$ is a unit vector in the direction of $\mathbf{r}-\mathbf{r}_{0,i}$, $\mathbf{m}_i$ is the unit vector in the direction of the $i$th dipole moment, $R^{\text{dip}}_i$ is the characteristic length scale, and $M_i$ is the magnitude.
The background solar dipole ($i = 0$) is located at $\mathbf{r}_{0,0} = [0,0,0]$, pointing north ($\mathbf{m}_0 = \hat{z}$) with $M_0 = 10$~G and $R_0^{\text{dip}} = R_s$. The active region is comprised of two sets of dipoles oriented in the $\hat{\theta}$ direction. 
For $i \in [1,21]$, $R_i^{\text{dip}} = 1.5\times 10^{10}$~cm, $M_i = 0.5$~G, and
\begin{equation*}
\mathbf{r}_{0,i} = \left[0.928571 R_s,0.5\pi,0.125\pi \left( \frac{i-11}{10} \right) \right].
\end{equation*}
For $i \in [22,30]$, $R_i^{\text{dip}} = 3\times 10^{10}$~cm, $M_i= 0.6$~G, and
\begin{equation*}
\mathbf{r}_{0,i} = \left[0.928571 R_s,0.5\pi,0.09375\pi \left( \frac{i-26}{4} \right) \right].
\end{equation*}
The distribution of dipoles is adapted from \citep{lynch08a}, to which we add a set of dipoles with larger $R_i^{\text{dip}}$ to broaden the active region in latitude.

\newpage
\bibliography{bibliography.bib}
\bibliographystyle{aasjournal}



\end{document}